\documentclass[twocolumn,showpacs,preprintnumbers,amsmath,amssymb]{revtex4}

\usepackage{amsfonts}

\usepackage{graphicx}
\usepackage{dcolumn}
\usepackage{bm}

\begin{document}


\title{Correlations in Grover search}

\author{Jian Cui }
\email{cuijian@aphy.iphy.ac.cn}
 \author{Heng Fan }%
 \email{hfan@aphy.iphy.ac.cn}
\affiliation{%
Institute of Physics, Chinese Academy of Sciences, Beijing 100190,
China
}%

\date{\today}

\begin{abstract}
Grover search is a well-known quantum algorithm that outperforms any
classical search algorithm. It is known that quantum correlations
such as entanglement are responsible for the power of some quantum
information protocols. But entanglement is not the only kind of
quantum correlations. Other quantum correlations such as quantum
discord are also useful to capture some important properties of the
nonclassical correlation. Also there is no well accepted and clear
distinction between quantum correlations and classical correlations.
In this paper, we systematically investigate several kinds of
correlations including both quantum and classical in the whole
process of Grover search algorithm. These correlations are the
concurrence, entanglement of formation, quantum discord, classical
correlation and mutual information. The behaviors of quantum
discord, classical correlation and mutual information are almost the
same while the concurrence is different in qubit-qubit case. For
qubit partition $1:n$ case, the behaviors of all correlations are
qualitative the same.  When the search is over, all kinds of
correlations are zero, we argue that this is necessary for the final
step in the search.

\end{abstract}

\pacs{03.65.Ud, 03.67.-a, 89.70.+c, 03.67.Lx}
\maketitle

\section{\label{sec:level2}Introduction}
It is known that quantum computing has great advantages over
classical ones by several quantum algorithms, e.g., the Grover
search algorithm \cite{r10} and Shor algorithm \cite{EJ}. The Grover
search algorithm provides a quadratic temporal speedup over the best
classical search algorithm when they both require the same spatial
resources to perform the same search task. It is believed that the
outstanding performance of a quantum computer comes from the quantum
phenomena such as quantum correlations, superposition, interference
etc. in its resources qubits. Quantum correlation, especially the
entanglement is one of the crucial issues in quantum information
theory and has been studied extensively \cite{HHHH}. It is clear
that quantum entanglement is essential in such tasks like quantum
teleportation \cite{BBCJPW}, superdense coding \cite{BW},
entanglement assisted classical capacity of the quantum channel
\cite{BSST}, etc. It is also believed that quantum entanglement is
necessary for Grover search \cite{r10} and Shor algorithm \cite{EJ}
though the role of entanglement is not as clear as for other quantum
information tasks like teleportation. Some properties of
entanglement in Grover search have been studied
\cite{Pati,Orus1,Orus2}. On the other hand, quantum entanglement may
not be necessary for a model of quantum information processing
introduced by Knill and Laflamme in Ref.\cite{r9}. Still such a
device can outperform its classical counterpart. Thus other quantum
correlations different from entanglement are necessary in describing
such a model. Ollivier and Zurek have recently defined the quantum
discord to measure the quantum correlations \cite{r1}. Datta {\it et
al} then applied quantum discord to characterize the correlations
present in the model introduced in Ref. \cite{r9}. They found that
while there is no entanglement between the control qubit and the
mixed ones, the quantum discord across this split is nonzero, see
Ref.\cite{r2}. Recently Lanyon {\it et al} implemented the above
model in an all-optical architecture, and experimentally observe the
generated correlations, see Ref.\cite{r3}.

Motivated by the above fact that in certain algorithm some kinds of
correlations such as quantum discord but not entanglement play a
fundamental role, we want to know whether this is true for Grover
search algorithm, by studying several well-known correlations as
well as entanglement in the process of search. In this paper, we
consider a quantum register consisting of n qubits. We adopt
concurrence as an entanglement measure and use quantum discord,
classical correlation and mutual information to quantify
correlation. To calculate entanglement or correlations, we have to
identify two subsystems, where two different methods are used. One
(i) is that naturally we divide n qubits into two subsystems
consisting of $k$ and $n-k$ qubits respectively. The other (ii) is
that we calculate a two qubit reduced density matrix and each
subsystem only contains one qubit.

When we use method (i) to divide the whole system into two
subsystems, we find that all correlations as well as entanglement
have qualitatively the same behaviors, See Fig(6-9). This suggests
that during Grover search the correlations in pure state can be
described well by any correlation measures, quantum or classical.
But when method (ii) is used, namely in a two-qubit reduced state,
the behaviors of correlations are different from that of
entanglement quantified by concurrence. Concurrence still firstly
increases to its maximum and then decreases to almost zero, but
other measurements of correlations repeat that routine for a second
time,see figure(2) and figure(3-5). We also find that the increasing
rate of success probability behaves the same way just as the
concurrence does. The concurrence and the increasing rate of success
probability get their maximal values almost at the same time.
Entanglement measured by concurrence acts as an indicator of the
increasing rate of success probability in Grover search. This
suggests in Grover search algorithm the place of entanglement cannot
be replaced by other correlations. The power of Grover search
depends on the ability to firstly increase the entanglement and then
eliminate it.

The paper is organized as follows. In Sec. II, we briefly review the
Grover search algorithm. In Sec. III, we introduce entanglement and
different kinds of correlations which will be uesd in this paper
including the mutual information, classical information and quantum
discord. In Sec. IV, we calculate the evolutions of the above
correlations during Grover search and show the results in several
figures. These results are analyzed in Sec.V and summarized in
Sec.VI.
\section{\label{sec:level6}Review of Grover search}

We briefly review the standard Grover search algorithm\cite{r4,r10}.
Suppose we have $n$-qubit register constructing a database of
dimension $N=2^n$. They are initialized in the pure
state$|0,\ldots,0\rangle$, and then subjected to local Hadamard
gates, $H^{\otimes n}$, where $H=(|0\rangle \langle 0|+|0\rangle
\langle 1|+|1\rangle \langle 0|-|1\rangle \langle 1|)/\sqrt{2}$. As
a result the register is in an equal superposition pure state
$|\psi\rangle=\frac{1}{N^{1/2}}\sum^{N-1}_{x=0}|x\rangle $. The
Grover search algorithm requires repeated routine (called
$iteration$), which can be expressed as
$G=(2|\psi\rangle\langle\psi|-I)O$, where $O$ is the oracle applied
in the algorithm. If the state is just the target state to search,
the oracle change the phase of the state by $\pi$, i.e.
$O|x\rangle=-|x\rangle$, when$|x\rangle$ is a state to search. If on
the contrary the state is not what we want, the oracle leaves it
invariant.

The $N$ states expressed by the $n$ qubits are divided into two
parts: the one belonging to the solution of the search, which is
expressed as $\sum^{'}|x\rangle$, and those which are not solutions
to the search which are expressed as $\sum^{''}|x\rangle $. The
normalized states are defined as
\begin{equation}
|m\rangle=\frac{1}{\sqrt{j}}\sum_{x}{} '|x \rangle
\end{equation}
\begin{equation}
|m^{\bot}\rangle =\frac{1}{\sqrt{N-j}}\sum_{x}{} ''|x \rangle,
\end{equation}
where $j $ is the total numbers of the target states. The two states
are orthogonal. For a simple case there is only 1 target state, i.e.
$j=1$. That is the case considered in this paper. It is easy to see
that the initial equal superposition state can be expressed in the $
|m^{\bot}\rangle$ and $|m\rangle$ bases as
\begin{equation}
|\psi\rangle=\sqrt{\frac{j}{N}}|m\rangle+\sqrt{\frac{N-j}{N}}|m^{\bot}\rangle.
\end{equation}
Next, we check the effect of the iteration. We set the two
orthonormal states $|m^{\bot}\rangle$ and $|m\rangle$ as two axes of
a rectangular coordinate system. The initial equal superposition
state $|\psi\rangle$ is a vector in the coordinate system. The
oracle operator reflects the vector $|\psi\rangle$ according to
$|m^{\bot}\rangle$. After that, the operator
$2|\psi\rangle\langle\psi|-I$ reflects the vector $O|\psi\rangle$
according to $|\psi\rangle$. These two steps together realize
turning the initial vector $|\psi\rangle$ towards the target vector
$|m\rangle$ by an angle of $\alpha$, if the angle between the vector
$|\psi\rangle$ and $ |m^{\bot}\rangle$ is $\frac{1}{2}\alpha$, see
Fig.1. Simple calculations yield that
$\alpha=\arccos(\frac{N-2j}{N})$. Therefore, by repeating the above
iteration routine the outcome state get more and more closer to the
target state. After $r$ times of iterations the result state
\begin{equation}
|\psi_{r}\rangle=\sin(\frac{2r+1}{2}\alpha)|m\rangle+\cos(\frac{2r+1}{2}\alpha)|m^{\bot}\rangle.
\end{equation}
The probability of success is
\begin{equation}
P=\sin^{2}\Big(\frac{2r+1}{2}\alpha\Big)
\end{equation}
which is an important parameter in Grover search. The best repeating
times to get the biggest probability of success is
\begin{equation}
R=CI\Big(\frac{\frac{\pi}{2}-\frac{\alpha}{2}}{\alpha}\Big)
\end{equation}
where $CI(x)$ denotes the integer closest to the real number $x$.

\begin{figure}
\includegraphics[height=6cm,width=\linewidth]{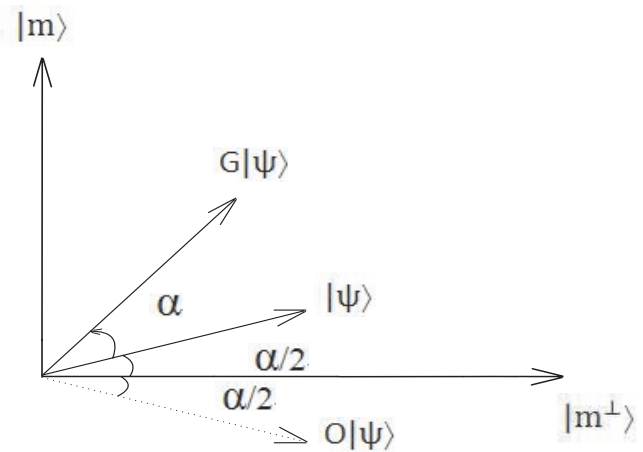}
\caption{\label{fig:0} This figure shows the role of iteration
played in the Grover search. $G=(2|\psi\rangle\langle\psi|-I)O$.
Firstly, $O$ reflects $|\psi \rangle$ according to $|m^\perp
\rangle$. Then $(2|\psi\rangle\langle\psi|-I$) reflects $O|\psi
\rangle$ according to $|\psi\rangle$. We can check that
$G|\psi\rangle+O|\psi\rangle=(G+O)|\psi\rangle=(2\langle\psi|O|\psi\rangle)|\psi\rangle$,
and what in the bracket is a C number. Therefore $|\psi\rangle$ is
the axis of the reflection form $O|\psi\rangle$ to $G|\psi\rangle$.
So one whole iteration turns the vector before iteration toward
$|m\rangle$ by an angle $\alpha$.}
\end{figure}

{\label{sec:level7}}\section {Quantum correlations }
 \subsection{entanglement}
Entanglement is viewed as a key resource in quantum computing. It is
believed to be responsible for the outstanding performances of
quantum computers compared with their classical counterparts in many
quantum information processing tasks, such as teleportation
\cite{BBCJPW}. There are many measurements of entanglement defined
from different considerations. Each measurement could capture
certain aspects of entanglement, but none of these measurements is
capable of involving all the features. Here we choose concurrence, a
well accepted entanglement measure, to investigate the behavior of
entanglement in the whole process of Grover search. Wootters has
defined concurrence for arbitrary state of two qubits in
Ref.\cite{r5}. For a two-qubit state $\rho$, we can first calculate
a relative matrix $\widetilde{\rho}=(\sigma_y \otimes
\sigma_y)\rho^*(\sigma_y\otimes\sigma_y)$, where $\sigma_y$ is the
Pauli matrix $\left(
\begin{array}{cccc}
0 &-i\\ i &0
\end{array}\right)$,
 and $\rho^*$ is the conjugation of $\rho$. Then the concurrence
\begin{equation}
  C(\rho)=max\left\{0,\lambda_1-\lambda_2-\lambda_3-\lambda_4\right\},
\end{equation}
and the $\lambda_{i}s$ are the square roots of the eigenvalues of
the matrix $\rho\widetilde{\rho}$ in decreasing order, i.e.
$\lambda_1>\lambda_2>\lambda_3>\lambda_4$. Fang {\it et al} have
calculated the concurrence between any two qubits in Grover search
by Wootters' formula, see Ref.\cite{r6}.

The above concurrence can be extended to the situation of higher
dimension pure bipartite state$|\psi\rangle$ \cite{r11,Fan,Gour}. We
will use this form to study the entanglement of Grover search. The
concurrence of $|\psi\rangle$ is defined as
\begin{equation}
C(|\psi\rangle)=\sqrt{\frac{d}{d-1}(1-Tr\rho^2_r)},
\end{equation}
where $\rho_r$ is the reduced density matrix obtained by tracing out
one of the two subsystems, and $d$ is the dimension of $\rho_r$. In
this paper, we will calculate the concurrence between any $k$ and
the other $n-k$ qubits with this formula and will also show this
result and other kinds of correlations in figures in the following.

\subsection{Quantum discord}
Quantum discord was first proposed by Ollivier and Zurek in\cite{r1}
as the difference between two expressions of mutual information
extended from classical to quantum system. Datta {\it et al} used
quantum discord to investigate a model which describes the power of
one qubit, see Ref.\cite{r7,r2,r9}. In fact, the quantum discord
that qualifies the quantum correlations can be viewed as the total
correlation subtracting the classical correlation. One version of
total correlation was defined by Groisman,Popescu and Winter in
\cite{GPW} in an operational way as the minimal amount of noise that
is required to erase all the correlations between the two systems.
They also showed that this definition of total correlation is equal
to the quantum mutual information. Quantum correlation and classical
correlation are generally involved together. To investigate
correlations in quantum algorithm, both classical and quantum
correlations are useful to capture some properties of the
correlation. Actually, when correlations, or other measurement data,
are sufficient to guarantee the existence of a certain amount of
quantum correlations in the system is a fundamental question in
particular while concerning about the measurements \cite{AP}.

In information theory, we know the total correlation between two
parties $A$ and $B$ is the mutual information denoted by $I(A,B)$,
see for example \cite{GPW}. For a quantum system
\begin{equation}
I(A,B)=S(\rho_A)+S(\rho_B)-S(\rho_{AB}),
\end{equation}
where $S(\rho)$ is the von Neumann entropy of $\rho$, $S(\rho)=-{\rm
Tr}(\rho {\rm log}\rho),$ and $\rho_A(\rho_B)$ is the reduced
density matrix of $\rho_{AB}$ by tracing out $B(A)$.

Classical correlation between $A$ and $B$ was defined by Henderson
and Vedral in Refs.\cite{r8,Vedral} as the maximum information we
can get from $A$ by measuring $B$.  Before measuring $B$, the
reduced density matrix of $A$ is $\rho_A$. Then we choose a complete
set of projectors $\{{\Pi_i}\}$ to measure the subsystem $B$,
corresponding to the outcome $i$ with the probability $p_i$. The
state of $A$ after the above measurement is
$\rho_{A|i}=\frac{Tr_B(\Pi_i\rho_{AB}\Pi_i)}{Tr_{AB}(\Pi_i\rho_{AB}\Pi_i)}$
, and $p_i=Tr_{AB}(\Pi_i\rho_{AB}\Pi_i)$. So the information of $A$
we can get by measuring $B$ is $S(\rho_A)-\sum_ip_iS(\rho_{A|i})$.
For a given density matrix $\rho_{AB}$, the above representation
depends on the choice of measurement, i.e. we can obtain different
results if we use different bases to apply the measurement. The
classical correlation measures the biggest amount of information,
that is
\begin{eqnarray}
C(A,B)=max_{\{\Pi_i\}}\{S(\rho_A)-\sum_ip_iS(\rho_{A|i})\}\nonumber\\
\mathcal=S(\rho_A)-min_{\{\Pi_i\}}\sum_ip_iS(\rho_{A|i}).
\end{eqnarray}
It can be checked easily that this definition of classical
correlation satisfies several conditions. These conditions include:
(i) $C=0$ for $\rho=\rho_A\otimes\rho_B$; (ii) $C$ is invariant
under local unitary transformations; (iii) $C$ is non-increasing
under local operations. (iv) $C=S(\rho_A)=S(\rho_B)$ for pure state,
see Ref.\cite{r8}.

Quantum discord expressed as $D$ is the difference between the total
correlation and the classical correlation, i.e.
\begin{eqnarray}
D(A,B)&=&I(A,B)-C(A,B)\nonumber\\
&=&min_{\{\Pi_i\}}\sum_ip_iS(\rho_{A|i})+(S(\rho_B)-S(\rho_{AB})).
\nonumber \\
\label{discord}
\end{eqnarray}
If we split the n-qubit system into one qubit slice and the other
$n-1$ qubits slice, and calculate the quantum discord between these
two parts, we can obtain a computable result theoretically. That is
because we can choose the one qubit slice as the part to be
measured, and have the bases of measurement parametrized by $\theta$
and $\phi$ in the form of
$\{\cos(\theta)|0\rangle+e^{i\phi}\sin\theta|1\rangle,
e^{-i\phi}\sin\theta|0\rangle-\cos\theta|1\rangle\}$. Therefore the
minimum according to $\{\Pi_i\}$ in equation (\ref{discord}) has
been changed into finding $\theta$ and $\phi$ to realize the
minimum. In the next section, we will use this method to calculate
the quantum discord between any one qubit and the other qubits .

{\label{sec:level8}}\section{Correlations in Grover search}
\subsection{Density matrix for the total system and the two qubits reduced density matrix}
We have already known the form of the state after $r$ times of
$iterations$ in Eq.(4)  in the $ |m^{\bot}\rangle$ and $|m\rangle$
bases. The density matrix is
\begin{widetext}
\begin{eqnarray}
\rho&=&\sin^2(\frac{2r+1}{2}\alpha)|m\rangle\langle
m|+\cos^2(\frac{2r+1}{2}\alpha)|m^\bot\rangle\langle m^\bot|
+\sin(\frac{2r+1}{2}\alpha)\cos(\frac{2r+1}{2}\alpha)(|m\rangle\langle
m^\bot|+|m^\bot\rangle\langle m|)\nonumber\\
&=&\sin^2\Big(\frac{2r+1}{2}\alpha\Big)\frac{1}{j}\sum_{i,k}{}'|i\rangle\langle
k|+\cos^2\Big(\frac{2r+1}{2}\alpha\Big)\frac{1}{N-j}\sum_{i,k}{}''|i\rangle\langle
k|\nonumber\\
&&+\sin\Big(\frac{2r+1}{2}\alpha\Big)\cos\Big(\frac{2r+1}{2}
\alpha\Big)\frac{1}{\sqrt{j(N-j)}}\Big(\sum_{i}{}'\sum_{k}{}''|i\rangle\langle
k|+\sum_{i}{}'\sum_{k}{}''|k\rangle\langle i|\Big)\nonumber\\
&=&a^2\frac{1}{j}\sum_{i,k}{}'|i\rangle\langle
k|+b^2\frac{1}{N-j}\sum_{i,k}{}''|i\rangle\langle
k|+ab\frac{1}{\sqrt{j}}\Big(\sum_{i}{}'\sum_{k}{}''|i\rangle\langle
k|+\sum_{i}{}'\sum_{k}{}''|k\rangle\langle i|\Big)
\end{eqnarray}
where the $\sum '$ stands for the sum of all the states belonging to
the search result,  and the $\sum ''$ means the sum of all the
states that are not what to search,
$a=\sin\Big(\frac{2r+1}{2}\alpha\Big)$ and
$b=\frac{1}{\sqrt{N-j}}\cos\Big(\frac{2r+1}{2}\alpha\Big)$ are
brought in to make the expression explicit. In the present work we
study the simplest case of having only one target state, i.e. $j=1$.
 Obviously, the above expression is in the computational
bases, and its matrix form is
\begin{eqnarray}
\bordermatrix{%
& & & & & & & &\cr
 &a^2&ab&ab&ab&ab&ab&ab&ab&\ldots\cr
 &ab&b^2&b^2&b^2&b^2&b^2&b^2&b^2&\ldots\cr
 &ab&b^2&b^2&b^2&b^2&b^2&b^2&b^2&\ldots\cr
 &ab&b^2&b^2&b^2&b^2&b^2&b^2&b^2&\ldots\cr
&ab&b^2&b^2&b^2&b^2&b^2&b^2&b^2&\ldots\cr
&ab&b^2&b^2&b^2&b^2&b^2&b^2&b^2&\ldots\cr
&ab&b^2&b^2&b^2&b^2&b^2&b^2&b^2&\ldots\cr
&ab&b^2&b^2&b^2&b^2&b^2&b^2&b^2&\ldots\cr
 &\vdots&\vdots&\vdots&\vdots&\vdots&\vdots&\vdots&\vdots&\ddots\cr
 }_{N\times N}
\end{eqnarray}
We can get the two-qubit reduced density matrix from the above
$n$-qubit one by tracing out any $n-2$ qubits. Mathematically the
result can be obtained in the following way: the above
$2^n\times2^n$ matrix is first divided into $2^{n-2}\times2^{n-2}$
parts symmetrically with every part a $4\times4$ matrix. Now the
initial matrix becomes a $2^{n-2}\times2^{n-2}$ matrix whose every
element is a $4\times4$ matrix. Then we sum the $2^{n-2}$ diagonal
elements up to get a new $4\times 4$ matrix,which is the reduced
density matrix for any two qubits. It takes the following form,
\begin{equation}
\begin{array}{cccc}
\rho_{2}=\left(\begin{array}{cccccccc}& a^2+(\frac{N}{4}-1)b^2
&ab+(\frac{N}{4}-1)b^2 &ab+(\frac{N}{4}-1)b^2
&ab+(\frac{N}{4}-1)b^2\\
\\
& ab+(\frac{N}{4}-1)b^2 &\frac{N}{4}b^2& \frac{N}{4}b^2& \frac{N}{4}b^2\\
\\
 & ab+(\frac{N}{4}-1)b^2 &\frac{N}{4}b^2 &\frac{N}{4}b^2 &\frac{N}{4}b^2\\
 \\
 &ab+(\frac{N}{4}-1)b^2 &\frac{N}{4}b^2 &\frac{N}{4}b^2 &\frac{N}{4}b^2\\
 \end{array} \right) .
\end{array}
\end{equation}
\end{widetext}
\subsection{Concurrence and other kinds of correlations in Grover search}
Based on this reduced density matrix, Fang {\it et al} \cite{r6}
used Wootters' formula mentioned above in equation (7) and
calculated the concurrence between any two qubit sites as follows,
\begin{eqnarray}
C_{1,1}&=&2\big|\cos(\theta_0-r\alpha)-\frac{1}{\sqrt{N-1}}\sin(\theta_0-r\alpha)\big|\times\nonumber\\
&&\frac{1}{\sqrt{N-1}}\sin(\theta_0-r\alpha).
\end{eqnarray}
This is the analytic pairwise entanglement in Grover search. The
evolution of the pairwise entanglement in Grover search algorithm is
calculated numerically and the result is shown in Fig.2 compared
with the probability of success in the search algorithm.

\begin{figure}
\includegraphics[height=8cm,width=\linewidth]{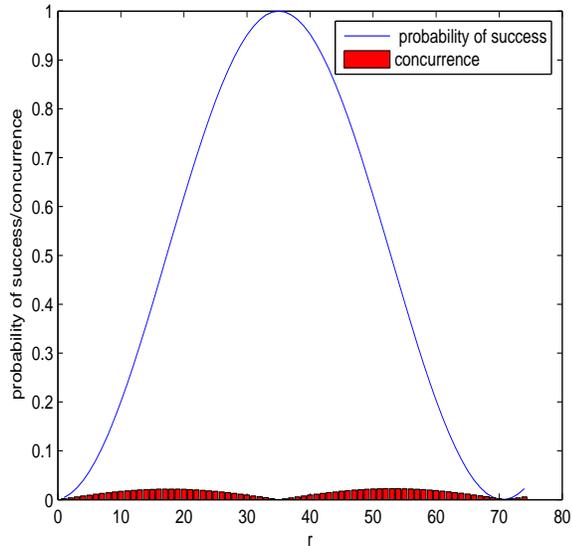}
\caption{\label{fig:1}(color online). Concurrence between any two
qubits for  N=2048 is obtained  using formula (15) based on the
reduced density matrix $\rho_2$ in equation (14). Similar result was
first obtained in \cite{r6} where N=256. }
\end{figure}


We can also use the two qubits reduced density matrix (14) to
calculate the mutual information, classical correlation and quantum
discord between any two qubits.  We first calculate the mutual
information using equation (9). The numerical results are presented
in Fig.3. In the calculation, $\rho_{AB}$ takes $\rho_2$, and
$\rho_A=\rho_B$ is the reduced density matrix by tracing out one
qubit from $\rho_2$.

Next, we will compute numerically the classical correlation
according to equation (10). Still $\rho_{AB}=\rho_2$ in Eq.(14) and
$\rho_A=\rho_B$ is the one qubit reduced density matrix. The main
task in this calculation is to find the minimum entropy of one qubit
after the measurement on another. The measurement bases are
parameterized by $\theta$ and $\phi$ in the form
$\{\cos(\theta)|0\rangle+e^{i\phi}\sin\theta|1\rangle,
e^{-i\phi}\sin\theta|0\rangle-\cos\theta|1\rangle\}$, where $\theta$
and $\phi$ both vary from $0$ to $2\pi$. For given $\theta$ and
$\phi$, thus a given measurement $\{{\Pi_i}\}$, we can obtain a
matrix
$\rho_{A|i}=\frac{Tr_B(\Pi_i\rho_{AB}\Pi_i)}{Tr_{AB}(\Pi_i\rho_{AB}\Pi_i)}$
with the probability $p_i=Tr_{AB}(\Pi_i\rho_{AB}\Pi_i)$
corresponding to the measurement's outcome $i$. With fixed $p_i$ and
$\rho_{A|i}$, we can calculate $\sum_ip_iS(\rho_{A|i})$. The aim in
computing the classical correlation is to find the minimum
$\sum_ip_iS(\rho_{A|i})$ depending on $\theta$ and $\phi$. We do it
numerically by choosing 256 values from $0$ to $2\pi$ for $\theta$
and $\phi$ respectively. For a density operator $\rho _2$, we can
find its classical correlation by optimizing the measurement
(finding optimal $\theta $ and $\phi $ ranging from 0 to $2\pi $).
The density operator $\rho _2$ varies in the Grover search
algorithm, the evolution of its classical correlation can thus be
calculated numerically. The results are presented in Fig.4.

We can find that quantum discord is the difference between the
mutual information $I$ and the classical correlation $C$, see
Eq.(11). The quantum discord can thus be obtained. The numerical
results are shown in Fig.5, where in all the above we take $N=2048$.

\begin{figure}
\includegraphics[height=7cm,width=\linewidth]{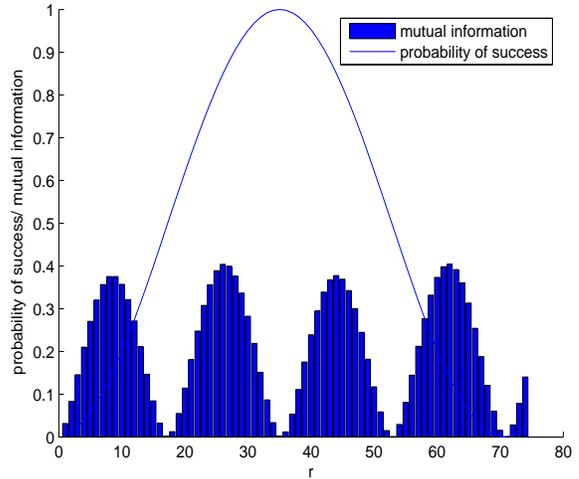}
\caption{\label{fig:5}(color online). Mutual information between any
two qubits for $N=2048$. The result is obtained numerically using
the formula in Eq.(9) based on the reduced density matrix in
Eq.(14)}
\end{figure}

\begin{figure}
\includegraphics[height=7cm,width=\linewidth]{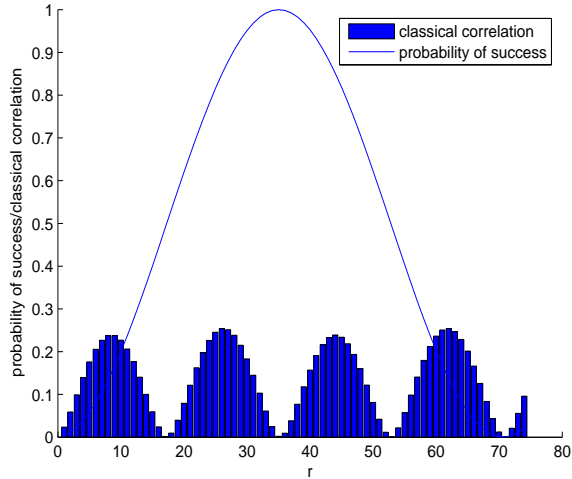}
\caption{\label{fig:6}(color online). Classical correlation between
any two qubits for $N=2048$. The result is obtained numerically from
Eq.(10). For each density operator, we have done the minimizing
procedure to obtain the classical correlation.}
\end{figure}

\begin{figure}
\includegraphics[height=8cm,width=\linewidth]{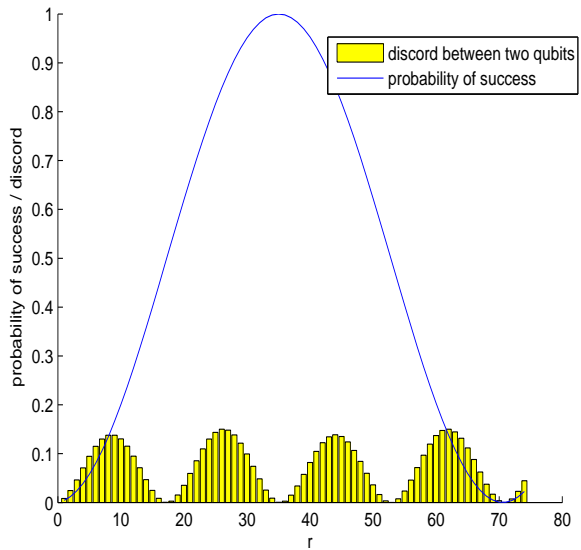}
\caption{\label{fig:2}(color online). Quantum discord between any
two qubits for $N=2048$. It is the difference between the mutual
information and classical correlation.}
\end{figure}

For multipartite system, besides the pairwise correlations between
any two qubits, other correlations are also of interest. In
particular, the pairwise entanglement sharing and other pairwise
correlations are monogamy \cite{GPW,KW,KBI,CKW,OV,OF}, when $n$
tends to infinity all of the pairwise correlations should vanish.
Therefore it should be interesting to view those correlations from a
different point. We will next study those correlations between any
one qubit and the other $n-1$ qubits. In this situation, we divide
the whole $n$ qubits into two parts: the $n-1$ qubits part $A$  and
the one qubit part $B$, where we choose $B$ as the part to be
measured when computing the classical correlation and quantum
discord.

The calculation is similar to the pairwise case. We first get the
$n-1$ qubits reduced density matrix $\rho_A$  and the one qubit
reduced density matrix $\rho_B$ from the whole density matrix
$\rho_n$ in Eq.(13). Then we can calculate the mutual information
using Eq.(9). Since part $B$ is one qubit, we can employ the
parameterized measurement bases
$\{\cos(\theta)|0\rangle+e^{i\phi}\sin\theta|1\rangle,
e^{-i\phi}\sin\theta|0\rangle-\cos\theta|1\rangle\}$, following the
same approach of minimizing the entropy after the measurement on
$B$,  the classical correlation and quantum discord can be found
numerically. The results are shown in Fig.6,7,8, where we set $n=8$.

\begin{figure}
\includegraphics[height=8cm,width=\linewidth]{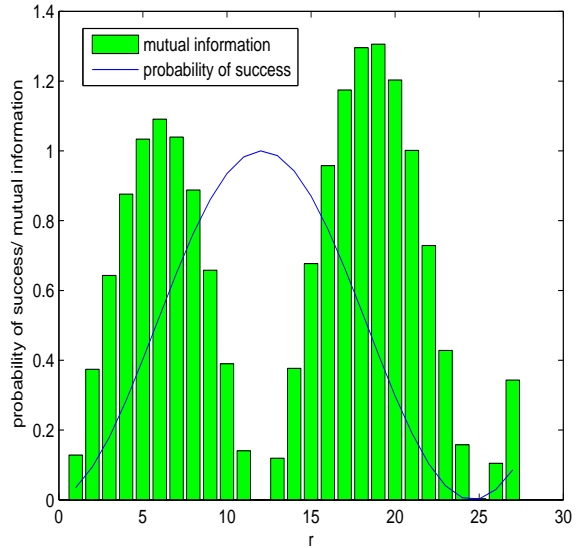}
\caption{\label{fig:11}(color online). Mutual information between
any one qubit and the other 7 qubits. We divide the whole $n$ qubits
into two parts: the $n-1$ qubits part  $A$  and the one qubit part
$B$, and use the formula in Eq.(9) to get the numerical result. Here
$n$ takes 8.}
\end{figure}

\begin{figure}
\includegraphics[height=8cm,width=\linewidth]{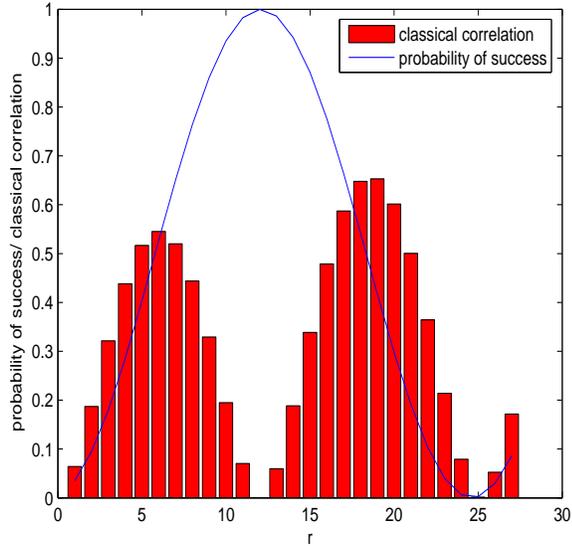}
\caption{\label{fig:10}(color online). Classical correlation between
any one qubit and the other 7 qubits. We notice that the behavior of
the classical correlation and the quantum discord are the same}
\end{figure}

\begin{figure}
\includegraphics[height=8cm,width=\linewidth]{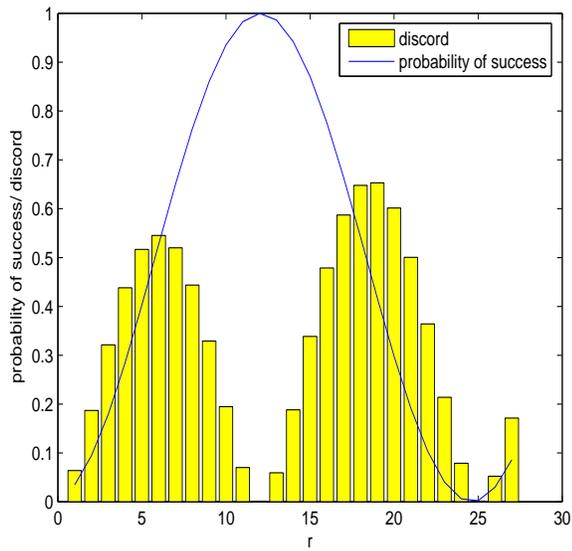}
\caption{\label{fig:9}(color online). Quantum discord between any
one qubit and the other 7 qubits. }
\end{figure}

For entanglement, we can also calculate the concurrence between any
$k$ and $n-k$ qubits. As can be seen from Eq.(4) that during the
whole process of Grover search the $n$-qubit register state is
always a pure state. So we can use Eq.(8) to calculate the
concurrence between any $k$ qubits and the other $n-k$ qubits,

\begin{eqnarray}
C_{k,n-k}&=&\left( \frac{2^k}{2^k-1} [1-(a^2+(\frac{N}{2^k}-1)b^2)^2
\right. \nonumber \\
&&-2(2^k-1)(ab+(\frac{N}{2^k}-1)b^2)^2 \nonumber \\
&&\left. -(1-2^{-k})^2N^2b^4] \right ) ^{1/2} .
\end{eqnarray}

For explicit, we show these results in Fig.9.

\begin{figure}
\includegraphics[height=8cm,width=8cm]{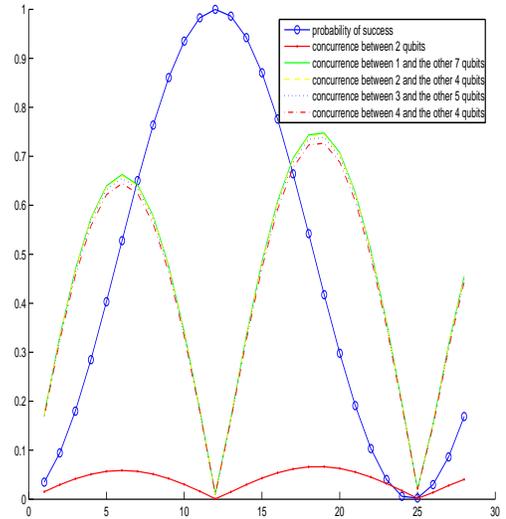}
\caption{\label{fig:7}(color online).  Concurrence between any $k$
and $n-k$ qubits, where $n=8$ and $k$ varies from 1 to 4. It is seen
from the graph that when $k$ varies the curve of concurrence doesn't
change a lot, which suggests that the entanglement between any two
parts is not sensitive to how you divide the whole register.}
\end{figure}

It is also of interest to compare entanglement and the mutual
information of any two qubits in the whole process. Here we use
entanglement of formation(EOF) as the entanglement measurement in
place of concurrence, since both mutual information and EOF are
defined by means of entropy. We find when EOF get its maximal mutual
information is minimal. The result is show in Fig.10.
\begin{figure}
\includegraphics[height=8cm,width=\linewidth]{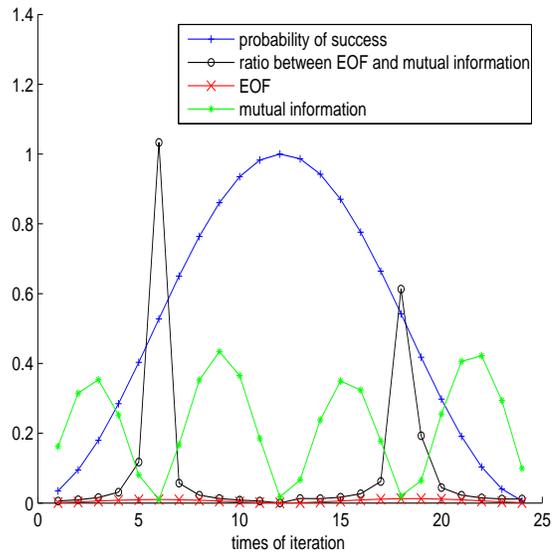}
\caption{\label{fig10}Comparison between EOF and mutual information
between any two qubits. n=8. }
\end{figure}
\subsection{The increasing rate of success probability vs entanglement}
The increasing rate is
\begin{equation}
\frac{\partial P}{\partial r}=\alpha sin((2r+1)\alpha).
\end{equation}
We find that the increasing rate has an interesting connection with
the concurrence between any two qubits in Eq.(15). Both of them
firstly increase with the iteration progress until get their summit
respectively and then begin to fall. Suppose they get their peak
point at $r1,r2$ respectively. We find that $r1,r2$ are connected.
From Eq.(17) we can calculate $r1$ directly. Let $(2r+1)\alpha =\pi
/2$ and remember that $r$ is an integer, so that
\begin{equation}
r1=CI\Big(\frac{1}{2}(\frac{\pi}{2\alpha}-1)\Big)
\end{equation}
To calculate $r2$ let $\frac{\partial C_{1,1}}{\partial r}=0$. We
can get
\begin{equation}
 r2=CI\Big(\frac{1}{2}(\frac{\pi}{2\alpha}-1.5)\Big)
\end{equation}
It can be seen directly form Eq.(18) and Eq.(19) that $r1-r2=0,1$.
We checked that both cases exist, e.g., when the total qubit number
$n=9,11,25,26,28,30,..., r1-r2=1,$ while $n$ takes other values
$r1-r2=0$. The relation between the increasing rate and entanglement
is shown in Fig.11.
\begin{figure}
\includegraphics[height=8cm,width=\linewidth]{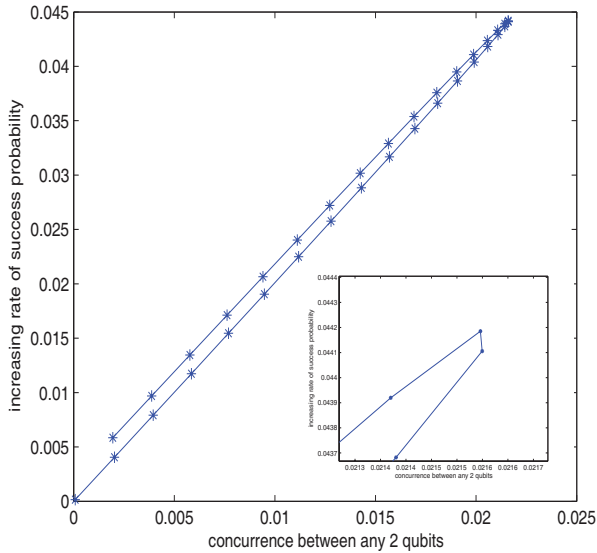}
\caption{\label{fig:11}The increasing rate of success probability vs
concurrence between any two qubits. Here we take $n=11.$ The part in
the vicinity of the peak point is enlarged at the corner.}
\end{figure}
\section{Analysis of the results and conclusions}
(a) We find that all these correlations mentioned above tend to
zeros near the point where the success probability of the search
runs to 1. This result indicates that when we fulfill the task of
searching, we have totally separated the target state. This can be
easily understood since when the search succeeds, what we got is the
final state (target state) which is a separable pure state. Thus
there is no correlations, quantum or classical. Since the target
state is separable, if there are any correlations existing, it means
that not yet the target state is obtained. Thus all correlations
being almost zeros is necessary for the final step in the search
algorithm. (b) We may also notice that in the initial state, all
correlations are also zeros. A naive guess may be that since the
target state and a database encoded in other $N-1$ ($N=2^n$) states
are superposed together in the initial state, the entanglement and
the correlations should be in a maximum point. Actually, since the
probable states are superposed together with same amplitude, the
initial state we prepared takes the form $|\psi \rangle =\frac
{1}{\sqrt{N}}\sum _{x=0}^N|x\rangle $. This is also a separable
state. For example, consider state $|\psi \rangle =\frac {1}{\sqrt
{d_1d_2...d_n}}\sum _{i_1i_2...i_n}|i_1i_2...i_n\rangle =(\frac
{1}{\sqrt {d_1}}\sum _{i_1}|i_1\rangle )...(\frac {1}{\sqrt
{d_n}}\sum _{i_n}|i_n\rangle )$, this is apparently a separable pure
state. Thus all correlations in the beginning of search are zeros.
(c) In the process of the search, we may find that the amplitude of
the target state becomes large monotonically, while the amplitudes
of other states are depressed. Thus the probability to find the
target state is enhanced in the process of the search until it
reaches the optimal point, and the probability to find other states
are negligible at that time. (d) The entanglement between any two
qubits quantified by concurrence firstly increases from zero to a
maximal point, then will decrease to zero. See figure(2). On the
other hand, the behaviors of classical correlation, quantum discord
and the mutual information between two qubits are different from the
behavior of the concurrence. After increasing and decreasing for the
first time, they repeat the routine for a second time. See
figure(3,4,5). When the concurrence reaches the maximal point, those
correlations become zeros. Our explanation is that at this case, the
correlations in the state are mainly entanglement, quantum discord
which also quantifies one property of the quantum correlations does
not exist at this point. This fact confirms the original claim in
Ref.\cite{r1} that quantum discord is a complementary quantity to
entanglement. (e) When investigating the total system whose state is
pure, the behaviors of all correlations between one and the other
$n-1$ qubits are actually the same. This suggests us that the pure
state correlations can be described by any of the correlation
measures. There is no qualitative difference between those measures.
(f) When the probability to find the target state is optimal, and
all correlations are almost zeros, at this time, if we continue the
search algorithm, all correlations will increase as presented in our
figures. And finally the state is expected to go back to the initial
state. (g) Entanglement is probably the reason for the increasing of
success probability in Grover search, i.e., the increasing rate of
success probability increases in accordance with entanglement, and
it get its maximum at the same time or immediately after the
entanglement approaches its summit. This result is another example
and further explanation of the argument by Shimoni, Shapira and
Biham in \cite{ssb} \cite{SSB}, where they applied Groverian
entanglement measure to characterize pure quantum state and argue
the entanglement is found to be correlated with the speedup achieved
by the quantum algorithm compared to classical algorithms. This also
explains why the power of the Grover search algorithm depends on the
ability to generate entanglement in the early stages of its
operation and on the ability to remove it when the target state is
approached\cite{SSB}.

{\label{sec:level9}} \section{summary}

In this work we have studied several correlations in the whole
process of Grover search and made a comparison among them. The
evolution results in the search algorithm obtained are quantities:
(i) the concurrence, entanglement of formation, quantum discord,
classical correlations and mutual information between any two
qubits; (ii) the concurrence between any $k$ qubits and the other
$n-k$ qubits; (iii) the quantum discord, classical correlation and
mutual information between any one qubit and the other $n-1$ qubits.
We have characterized the Grover search algorithm and showed the
results in figures. In particular in these figures we gave the
evolution of quantum discord in the whole process of Grover search
which had never been obtained before to our knowledge. We also argue
that entanglement measured by concurrence works as the indicator of
the increasing rate of the success probability.

The role of different kinds of correlations in quantum information
processing tasks is an interesting question. We systematically
studied evolution of several correlations in Grover search. It will
also be interesting to study correlations in other quantum
algorithms.

{\label{sec:level13}} \section{Acknowledgements}

One of the authors, H.F. acknowledges the support by "Bairen"
program, NSFC grant (10674162) and "973" program (2006CB921107).


\begin{thebibliography}{99}


\bibitem{r10}L. K. Grover, Phys. Rev. Lett.\textbf{79},325 (1997).
\bibitem{EJ}A. Ekert, and R. Jozsa, Rev. Mod. Phys. {\bf 68}, 733
(1996).
\bibitem{HHHH}R. Horodecki, P. Horodecki,  M. Horodecki,  K.
Horodecki, Rev. Mod. Phys. (2009).
\bibitem{BBCJPW}C. H. Bennett, G. Brassard, C. Crepeau,
R. Jozsa, A. Peres, W. K. Wootters, Phys.Rev.Lett. {\bf 70},
1895(1993).
\bibitem{BW} C. H. Bennett, S. J. Wiesner, Phys. Rev. Lett. {\bf
69}£¬ 2881 (1992).
\bibitem{BSST} C. H. Bennett, P. W. Shor, J. A. Smolin,
and A. V. Thapliyal, Phys. Rev. Lett. {\bf 83}, 3081 (1999).
\bibitem{Pati}S. L. Braunstein and A. K. Pati, Quant. Info. Comput.
{\bf 2}, 399 (2002).
\bibitem{Orus1}R. Orus and J. I. Latorre, Phys. Rev. A {\bf 69},
052308 (2004).
\bibitem{Orus2}R. Orus, J. I. Latorre, and M. A. Martin-Delgado,
Eur. Phys. J. D {\bf 29}, 119 (2004).
\bibitem{r9}E. Knill and R. Laflamme, Phys. Rev. Lett. \textbf{81},5672 (1998)



\bibitem{r1}H. Ollivier and W. H. Zurek, Phys. Rev. Lett. \textbf{88}, 017901 (2002)
\bibitem{r2}A. Datta, A. Shaji and C. M. Caves, Phys. Rev. Lett. \textbf{100},
050502 (2008)
\bibitem{r3}B. P. Lanyon, M. Barbieri,M. P. Almeida, and A. G. White, Phys. Rev. Lett. \textbf{101}
, 200501 (2008)
\bibitem{r4}M. A. Nielsen and I. L. Chuang: \emph{Quantum
Computation and Quantum information}, Cambridge University Press,
Cambridge 2000.
\bibitem{r5}W. K. Wootters, Phys. Rev. Lett. \textbf{80}, 2245 (1998)
\bibitem{r6}Y. Y. Fang, D. Kaszlikowski, C. Chin, K. Tay, L. C. Kwek and C. H. Oh,
Phys. Lett. A \textbf{345}, 265 (2005)
\bibitem{r11}P. Rungta, V. Buzek, C. M. Caves, M. Hillery,
and G. J. Milburn, Phys. Rev. A {\bf 64}, 042315 (2001).
\bibitem{Fan}H. Fan, K. Matsumoto, and H. Imai, J. Phys. A {\bf 36},
4151 (2003).
\bibitem{Gour}G. Gour, Phys. Rev. A. \textbf{71},
012318 (2005).
\bibitem{r7}A. Datta, S. T. Flammia and C. M. Caves,  Phys. Rev. A.
\textbf{72}, 042316 (2005)





\bibitem{GPW} B. Groisman, S. Popescu, and A. Winter, Phys. Rev. A
{\bf 72}, 032317 (2005).
\bibitem{AP}K. M. R. Audenaert and M. B. Plenio, New J. Phys.
{\bf 8}, 266 (2006).
\bibitem{r8}L. Henderson, and V. Vedral, J. Phys. A. \textbf{34},6899
(2001).
\bibitem{Vedral}L. Henderson, and V. Vedral, Phys. Rev. Lett. {\bf
84}, 2263 (2000).
\bibitem{KBI} M. Koashi, V. Buzek, and N. Imoto, Phys. Rev. A {\bf
62} , (2000).
\bibitem{CKW}V. Coffman, J. Kundu, and W. K. Wootters, Phys. Rev. A
{\bf 61}, 052306 (2000).
\bibitem{KW} M. Koashi, and A. Winter, Phys. Rev. A {\bf 69},
022309 (2004).
\bibitem{OV}T. J. Osborne and F. Verstraete, Phys. Rev. Lett. {\bf
96}, 220503 (2006).
\bibitem{OF}Y. C. Ou, and H. Fan, Phys. Rev. A {\bf 75}, 062308
(2007).
\bibitem{ssb}Y.Shimoni, D.Shapira and O. Biham, Phys. Rev. A{\bf
72},062308(2005)
\bibitem{SSB}Y.Shimoni, D.Shapira and O. Biham, Phys. Rev. A{\bf
69},062303(2004)



\end{thebibliography}
\end{document}